# MEMS Cantilever Based Frequency Doublers

**Joydeep Basu**[1], **Tarun Kanti Bhattacharyya**[2]

Department of Electronics and Electrical Communication Engineering

Indian Institute of Technology Kharagpur, Kharagpur 721302, India

Email: [1]joydeepkgp [at] yahoo.com, [2]tkb [at] ece.iitkgp.ernet.in

**Abstract:** Microelectromechanical system (MEMS) based on-chip resonators offer great potential for high frequency signal processing circuits like reference oscillators and filters. This is due to their exceptional features like small size, large frequency-quality factor product, integrability with CMOS ICs, low power consumption, low cost batch fabrication etc. A capacitively transduced cantilever beam resonator is one such popular MEMS resonator topology. In this letter, the inherent square-law nonlinearity of the voltage-to-force transfer function of a cantilever resonator's capacitive transducer has been employed for the realization of frequency doubling effect. Using this concept, frequency doubling of input signals of 500 kHz to 1 MHz, and 227.5 kHz to 455 kHz has been experimentally demonstrated for two cantilever beams of length 51.75 μm and 76.75 μm respectively. The MEMS cantilevers have been fabricated with polysilicon using the PolyMUMPs surface micromachining process, and their testing has been performed using Laser Doppler Vibrometry. The test results obtained are in reasonable compliance with the analytical and CoventorWare finite-element simulation results. The high efficiency demonstrated by the cantilever frequency doubler makes it a promising choice for signal generation at high frequencies.

# 1. Introduction

Over the past few years, microelectromechanical system (MEMS) based resonators have emerged as an attractive alternative for quartz crystal resonators for sensing and high frequency signal processing applications. Some of their promising aspects include quality factor (Q) values close to that for quartz in both vacuum as well as in air and operating frequencies in the very-high frequency (VHF) and ultra-high frequency (UHF) ranges, low power consumption, compact size, and the feasibility of monolithic integration and fabrication using low-cost CMOS compatible processes (Nguyen, 2007; Basu and Bhattacharyya, 2011b). Micromechanical cantilevers vibrating in flexural modes are one of the most popular MEMS resonators used in various applications of micro and nanotechnologies of the present age (Lin et al., 2004; Li, Tang and Roukes, 2007; Chakraborty and Bhattacharyya, 2010). The simplicity of the structure and its operation, the ease of fabrication over a wide range of dimensional variations, the ease of excitation and resonance measurements, and the extensively researched analytical background contribute towards this popularity.

Frequency multiplication has been an indispensable part of radio frequency (RF) communication technology. The central principle followed for achieving this lies in the utilization of a nonlinear element for generating harmonics of an input signal of a particular frequency. Hence, we can generate signals at higher frequencies using this technique. Mainly diode and field-effect transistor (FET) based frequency multipliers have seen rapid development in the last few decades, with the best Schottky diode frequency multiplier of today being able to generate signals with frequencies of the order of 1 THz (Faber, Chramiec and Adamski, 1995; Camargo, 1998; Golio, 2002; Ward et al., 2004). However, all these have their intrinsic limitations, with a typical frequency doubler losing about 70% of the input signal power to heat (Wang et al., 2009). Recently, graphene based FETs have been employed for frequency doubling with more than 90% converting efficiency (Wang et al., 2009; Wang et al., 2010). But, a maximum generated frequency of only up to 400 kHz has been reported. Therefore, it can said that the development of frequency doublers is still a topic of interest, as we need such devices for a plethora of applications like wireless communication, radio astronomy etc. This letter demonstrates for the first time the potential of MEMS cantilevers for usage in frequency doubling application. Due to the low-loss attribute (or equivalently, a high Q) offered by MEMS devices, improved circuit performance can be expected from a micromechanical frequency doubler than semiconductor devices utilized for the same task (Rebeiz, 2003). The concept proposed in this work can be further extended by using bulk-mode MEMS resonator geometries, like a circular-disk resonator (Wang, Ren and Nguyen, 2004; Basu and Bhattacharyya, 2011a) for the generation of frequencies up to the UHF range.

## 2. Theory

An electric-field driven micromechanical cantilever resonator is one of the most fundamental and widely studied structures in MEMS, which can provide a high Q and narrow bandpass filtering function. The diagram of a simple cantilever beam fixed at one end is given in Figure 1.

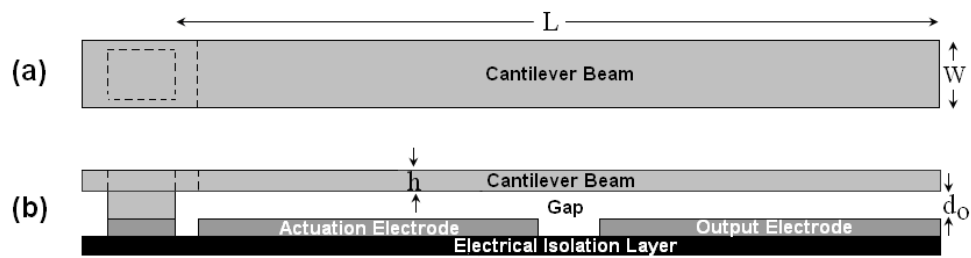

Fig. 1. Top view (a) and side view (b) of the cantilever resonator.

The instantaneous deflection under undamped free vibration of the beam of length L, rectangular cross-section with width W and thickness h, as shown in the figure, is governed by the Euler-Bernoulli equation (Bao, 2005):

$$EI\frac{\partial^4 y}{\partial x^4} + \rho A \frac{\partial^2 y}{\partial t^2} = 0 \tag{1}$$

Here y(x,t) is the out-of-plane deflection of the beam at a distance x from the clamped-end at a time instant t, ρ and E are the density and Young's modulus of the beam material respectively, A is the cross-sectional area, and I is the area moment of inertia of the beam. Solving this equation, the natural frequency of the beam for $n^{th}$ vibration mode can be determined as (Bao, 2005; Chakraborty and Bhattacharyya, 2010):

$$f_n = \frac{1}{2\pi}\left(\frac{k_n^4}{12}\right)^{0.5}\left(\frac{h}{L^2}\right)\sqrt{\frac{E}{\rho}} \tag{2}$$

Also, $k_n$ is equal to 1.875 for the first resonance mode of the beam (i.e., for n = 1). Before explaining its frequency doubling operation, we need to start with the concept of a cantilever being used as a resonator for which a dc-bias voltage $V_{dc}$ should be applied to the beam and an ac-excitation $v_i$ should be applied to the actuation (or, input) electrode. Thus, a resultant potential difference of ($V_{dc} - v_i$) is produced across the input-transducer comprising of the beam and input electrode. Due to this applied voltage, an electrostatic force ($F_e$) is generated between this electrode and the beam having both constant ($F_0$) and time varying ($F_i$) components, the expression of which can be derived using the energy method (Senturia, 2001):

$$F_e = \frac{\partial}{\partial y}\left(\frac{1}{2}C_i(V_{dc}-v_i)^2\right) = \frac{1}{2}(V_{dc}-v_i)^2\left(\frac{\partial C_i}{\partial y}\right) = F_0 + F_i \tag{3}$$

where $C_i(y)$ denotes the input-transducer capacitance. The time varying force $F_i$ is given in Equation (4), where only the dominant term (for $V_{dc} \gg v_i$) has been retained.

$$F_i \cong -V_{dc}v_i\left(\frac{\partial C_i}{\partial y}\right) = -V_{dc}v_i\frac{\partial}{\partial y}\left[\frac{\varepsilon A_i}{d_0}\left(1-\frac{y}{d_0}\right)^{-1}\right] \cong -V_{dc}v_i\left(\frac{\varepsilon A_i}{d_0^2}\right) \tag{4}$$

Here ε is the permittivity of the medium, $d_0$ is the static electrode-to-beam gap spacing, and $A_i$ is the input transducer area. Hence, this force acting on the beam has the same frequency as that of input $v_i$. Now, when the frequency of $v_i$ approaches the beam's resonant frequency ($\omega_1 = 2\pi f_1$), the beam begins to vibrate in resonance in a direction perpendicular to the substrate, creating a dc-biased time varying capacitor $C_o(t)$ between the beam and the output electrode. As given in Equation (5), $C_o(t)$ has a fixed as well as a sinusoidally varying component.

$$C_o(t) = C_{fix} + C_{var}\sin(\omega t) \quad (5)$$

For all input frequencies away from the resonance frequency, $C_o \approx C_{fix}$, But, at resonance, large deflections of the beam occur with $C_{var}$ assuming a significant value. Thus, in this condition (i.e., for $\omega = \omega_1$), an output motional current $i_o$ is produced at the output transducer as given by:

$$i_o = V_{dc}\frac{\partial C_o}{\partial t} = V_{dc}C_{var}\omega\cos(\omega t) \quad (6)$$

having a frequency of $\omega_1$. This frequency filtering function justifies the operation as a resonator. The output current can be directed to a load resistor $R_L$ which provides the proper termination impedance for the resonator.

Now, for using this cantilever as a frequency doubler rather than a resonator, we need to apply the dc potential ($V_{dc}$) also in series with the ac input excitation ($v_i$) to the input-electrode, as illustrated in Figure 2(a).

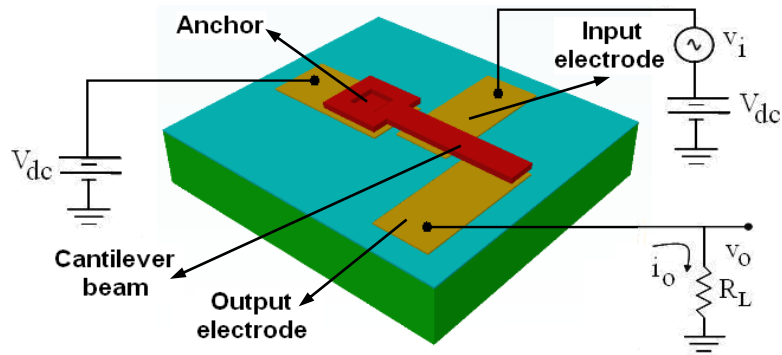

Fig. 2(a). Schematic diagram of the cantilever-based frequency doubler.

This is due to the fact that the electrostatic force $F_e$ produced on the beam is proportional to the square of the voltage-difference between the beam and the input-electrode (from Equation (3)) which is here equal to $v_i$. Hence, if the frequency of the sinusoidal $v_i$ is say, $f_i$, then the force $F_e$ acts at a frequency of $2f_i$. Now, if the cantilever is selected to have a resonant frequency ($f_1$) of the same $2f_i$, then the beam vibrates at resonance and we get a sinusoidal motional current $i_o$ with a frequency of $2f_i$ (using Equation (6)) in the load connected at the output port. Taking $A_o$ and $y(t)$ as the output-transducer overlap area and vertical deflection of the beam from the position of rest respectively, we have:

$$C_o(t) = \frac{\varepsilon A_o}{d_0 - y(t)} = \frac{\varepsilon A_o}{d_0(1 - y(t)/d_0)} \cong \frac{\varepsilon A_o}{d_0}\left(1 + \frac{y(t)}{d_0}\right) \tag{7}$$

If we assume that the sinusoidal $y(t)$ is given by $y_0 \sin[2\pi(2f_i)t]$, then $i_o$ has the following expression:

$$i_o = V_{dc}\frac{\partial C_o}{\partial t} = V_{dc}\frac{\varepsilon A_o y_0}{d_0^2}(4\pi f_i)\cos[2\pi(2f_i)t] \tag{8}$$

Hence, the output current has twice the frequency of the input signal.

## 3. Design, Simulation and Fabrication

For verifying the concept, two polysilicon cantilever beams of 2 μm thickness have been designed and fabricated using the PolyMUMPs surface micromachining technology (Carter et al., 2005), having lengths of 51.75 and 76.75 μm respectively. These are intended to have the first resonance mode at 1 MHz and 455 kHz correspondingly (using Equation (2)); and hence, can double an input signal frequency of 500 kHz to 1 MHz, and 227.5 kHz to 455 kHz respectively. The width has been taken as 10 μm. The beam has been realized using the Poly1 structural layer of polysilicon having a thickness of 2.0 μm and deposited by LPCVD (Low Pressure Chemical Vapor Deposition). The bottom electrodes have been patterned in the 500 nm thick Poly0 layer of polysilicon. The resulting electrode-to-beam air-gap is 2 μm. Electrical connections from the pads to the beam and I/O-electrodes have also been realized using the Poly0 layer. The two cantilever beam resonators have been simulated in CoventorWare Finite-Element Method (FEM) based simulation software (Basu et al., 2011c; Coventor, 2008), and modal analyses have been performed to determine the mode shapes and the corresponding modal frequencies. The first vibration mode for the 76.75 µm long beam is at 455 kHz, the simulated modal shape for which is shown in Figure 2(b).

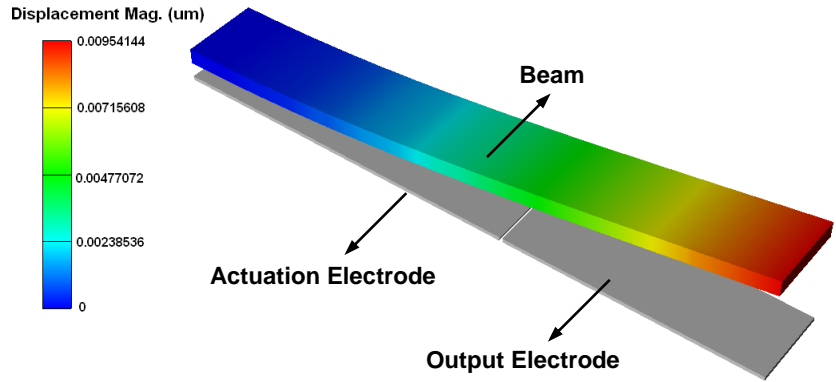

Fig. 2(b). First mode shape of the 76.75 µm long cantilever beam at 455 kHz.

Field-emission Scanning Electron Microscope (FESEM) image of a fabricated cantilever-based frequency doubling device is provided in Figure 3 which also illustrates the various material layers utilized in fabrication of the structure.

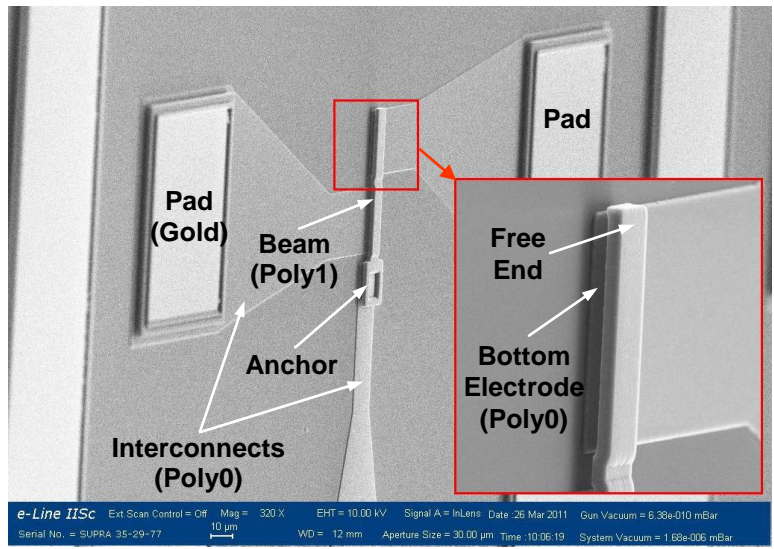

Fig. 3. SEM image of the 76.75 µm long cantilever frequency doubler. The inset shows a close-up view of the tip of the beam.

## 4. Characterization

The two cantilever resonator based frequency doubling devices have been electro-mechanically characterized at the wafer level using Laser Doppler Vibrometry (LDV) technique by which non-contact vibration measurements of a surface can be made in real time (Pratap and Pandey, 2007; Polytec, 2012). LDV uses the principle of *Doppler frequency shift* of back scattered laser light from a moving surface for measuring the component of velocity which lies along the axis of the laser beam. A Micro Scanning Laser Vibrometer (model: Polytec MSV 300) has been utilized for characterizing the frequency response of the designed cantilevers. The electrical excitation signal generated by the internal signal generator of the LDV system has been applied between the beam and any one of the two bottom electrodes to excite out-of-plane modes in the cantilever. The verification of the frequency doubling effect has been separately done for each of the cantilevers in two consecutive steps, by the application of two different kinds of signals as described below:

- In the first step, a *periodic chirp* input signal having a uniform distribution of energy across a frequency range from 0 to 2 MHz have been used for exciting the cantilever. This helps us in finding the resonant frequencies and corresponding modal shapes of the flexural vibration modes within this frequency range.
- In the second step, a *sinusoidal* voltage of frequency equal to half of the first resonance mode frequency ($f_1$, determined in the previous step) has been applied as the excitation signal. This causes the cantilever beam to vibrate at the first resonance mode. In this way, out-of-plane deflection has been excited in the beam at a frequency which is twice the frequency of the input voltage, which confirms the frequency doubling effect of the device.

The resonance frequency ($f_1$), quality factor (Q), and the damping ratio ($\zeta$) can be obtained from either of the vibration velocity/displacement/voltage responses as a function of the frequency of the applied signal. The vibration velocity spectrum for the two cantilever beam resonators subjected to periodic chirp input are shown in Figure 4(a) and (b).

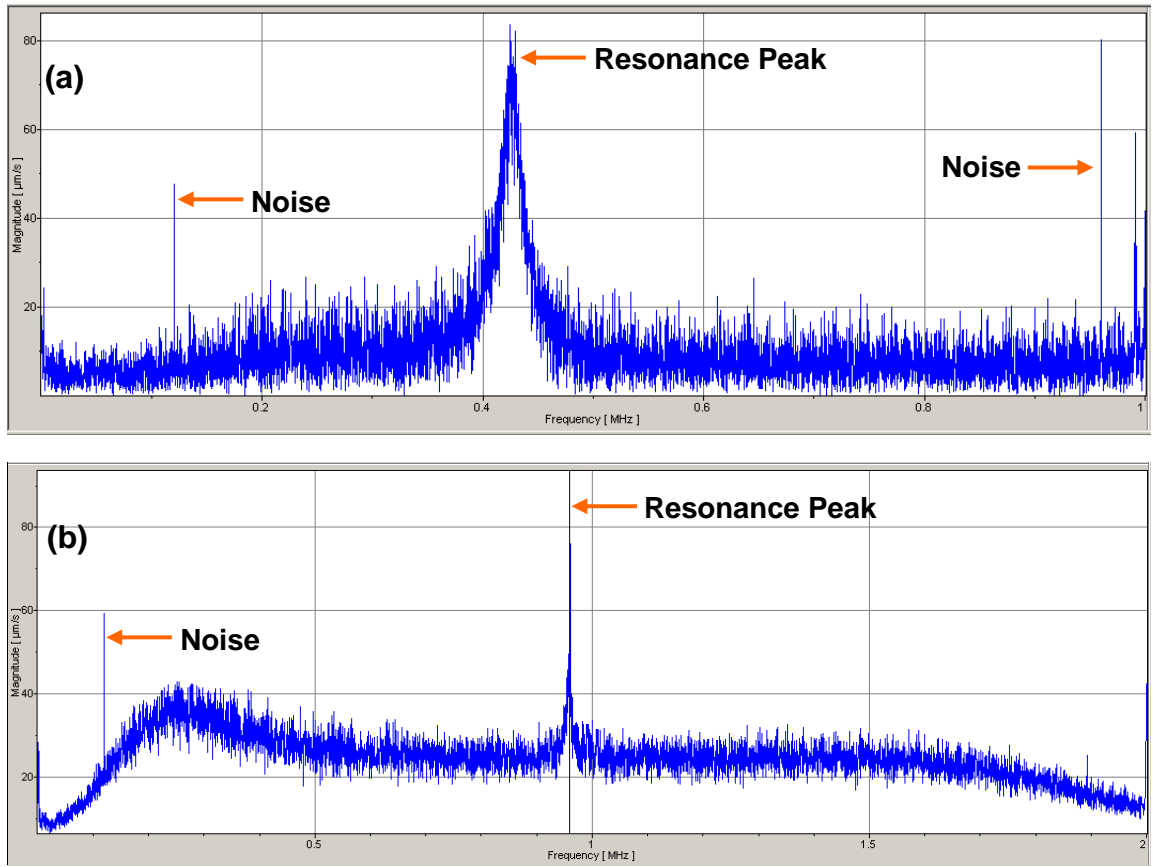

Fig. 4. Velocity spectrums of the 76.75 μm (a) and 51.75 μm (b) long cantilever beams obtained from LDV.

The measurements have been repeated for various values of the peak-amplitude of input $v_i$ (up to 10 V). The resonance frequency can be detected as the frequency at which the peak of velocity (or, displacement) amplitude occurs in the output spectrum. Hence, from the figures, it can be seen that the first mode of vibration is obtained approximately at 435 kHz and 960 kHz for the two cantilevers of length 76.75 μm and 51.75 μm respectively. A slight reduction of the fundamental resonance frequency is mainly due to squeezed-film air damping (Chakraborty and Bhattacharyya, 2010). All the measured spectrums have a spike at about 120 kHz due to system generated noise. The half-power or 3-dB bandwidth (Δf) can be obtained from the difference of frequencies at which the vibration velocity magnitude is 1/√2 times of the peak magnitude. Using the values of $f_1$ and $\Delta f$, the quality factor and the damping ratio (Pratap and Pandey, 2007) can be determined respectively from Equations (9) and (10).

$$Q = \frac{f_1}{\Delta f} \tag{9}$$

$$Q = \frac{1}{2\zeta\sqrt{1-\zeta^2}} \tag{10}$$

Thus, from the spectrum for the 76.75 μm long cantilever beam, the Q and ζ have been calculated as 40 and 0.0125 respectively.

The modal shape of the first vibration mode of each of the cantilevers has also been visualized using LDV. The one obtained for the 51.75 μm long cantilever beam is shown in Figure 5. Measured modal shape has been found to match with that obtained from FEM simulations. This also shows that the Poly1 beam has been properly released and is free at the end.

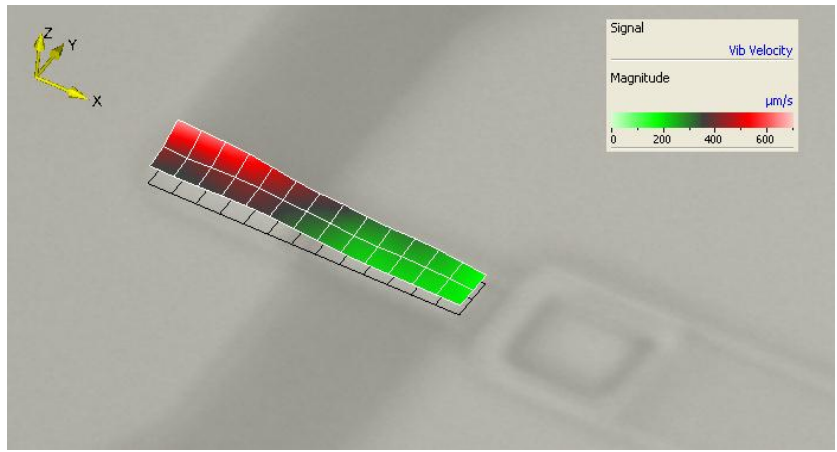

Fig. 5. The first mode vibrating beam shape for the 51.75 μm long cantilever beam obtained from LDV.

Next, a sinusoidal signal of frequency equal to half of the corresponding first mode resonance frequency has been used. Figure 6(a) shows the vibration velocity spectrum for the beam of length 76.75 μm. Here, the resonance peak has been obtained at 454 kHz by using an input of half of this frequency. Similarly, the output vibration voltage spectrum having a peak at 960 kHz for the second beam of length 51.75 μm is provided in Figure 6(b). This has been obtained by providing an input excitation at 500 kHz which is half the resonance frequency of this beam. Hence, the frequency doubling effect has been experimentally confirmed. All of the output spectrums have shown high spectral purity.

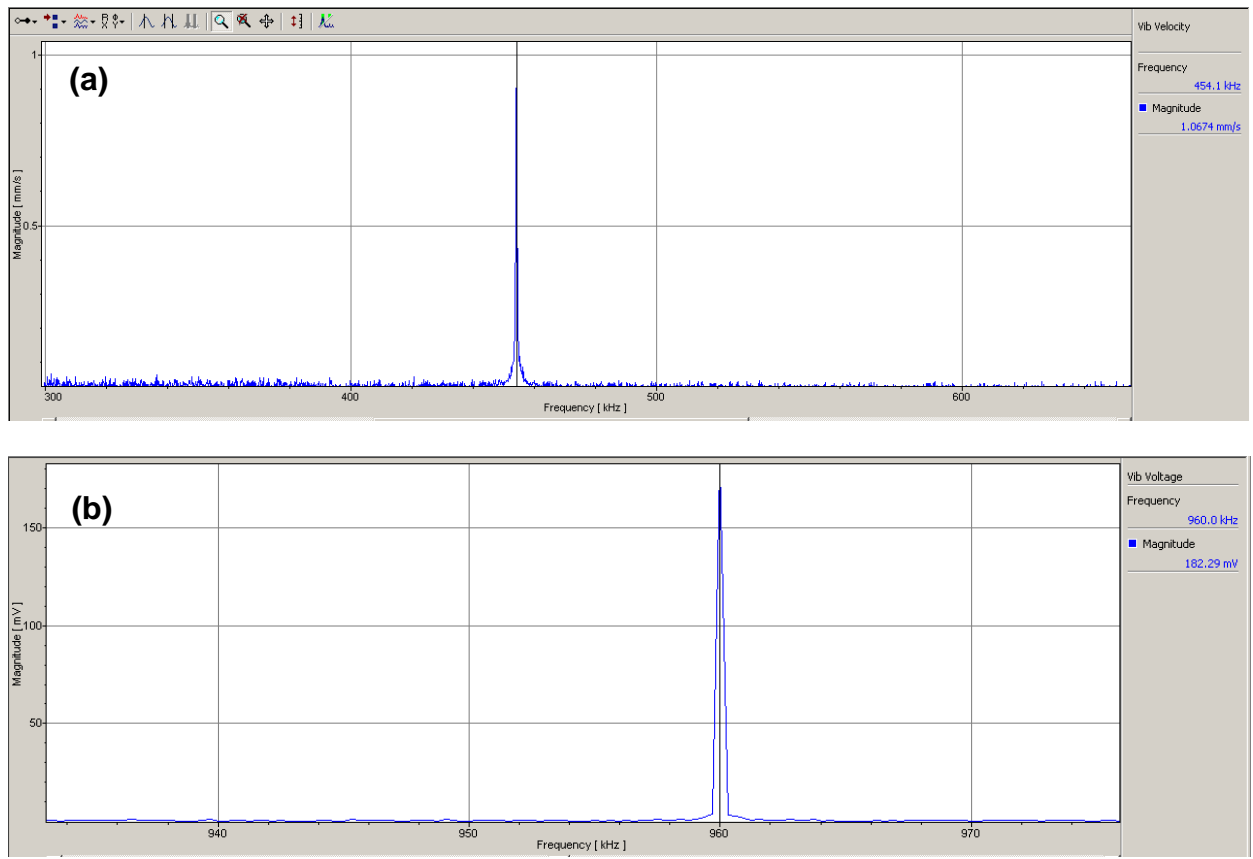

Fig. 6. Velocity spectrum of a 76.75 μm long cantilever (a) and vibration voltage spectrum of a 51.75 μm long cantilever (b) obtained from LDV for input at half the resonance frequency.

The same set of testing has also been repeated for similar devices but present on separate dies. Table 1 presents the experimental output frequencies obtained by application of sinusoidal input (at half frequency) to cantilevers on three such dies. Only a minor variation has been observed across the results.

**Table 1. Test results for identical cantilever beams on three separate dies.**

| Die no. | Measured output frequency of cantilever frequency doubler [in kHz] | |
|---|---|---|
| | First cantilever (Theoretical $f_1$ = 455) | Second cantilever (Theoretical $f_1$ = 1000) |
| 1 | 454 | 960 |
| 2 | 454 | 957 |
| 3 | 453 | 959 |

## 5. Conclusion

In this letter, MEMS cantilever beam fixed at one end has been utilized for realizing frequency doubler. The microstructures have been designed and fabricated using a surface micromachining process, and have also been finite-element simulated for their mechanical response. Theory and simulation have been found to match quite well. Further, the functionality of the frequency doubling device has been experimentally demonstrated using LDV technique for two fabricated cantilever beams. The testing has been repeated on devices on multiple dies. The test results have been found to be in reasonable compliance with the estimated results. The work reported here can be further continued with various prospects, including the detailed performance comparison of MEMS frequency doubler with comparable semiconductor versions, and the utilization of high-frequency bulk-mode resonator geometries for frequency doubling applications.

## Acknowledgement

We would like to thank National Programme on Micro and Smart Systems (NPMASS), Government of India for sponsoring project on RF-MEMS at IIT Kharagpur; and acknowledge Prof. Rudra Pratap and Prof. Navakanta Bhat of IISc Bangalore for providing us with the LDV and FESEM setups respectively.

# References

Bao, M. 2005. *Analysis and Design Principles of MEMS Devices*, 1st ed., Elsevier B. V., Amsterdam.

Basu, J. and Bhattacharyya, T.K. 2011a. "Comparative Analysis of a Variety of High-Q Capacitively Transduced Bulk-mode Microelectromechanical Resonator Geometries," *Microsystem Technologies*, 17(8):1361–1371.

Basu, J. and Bhattacharyya, T.K. 2011b. "Microelectromechanical Resonators for Radio Frequency Communication Applications," *Microsystem Technologies*, 17(10–11):1557–1580.

Basu, J., Chakraborty, S., Bhattacharya, A. and Bhattacharyya, T.K. 2011c. "A Comparative Study between a Micromechanical Cantilever Resonator and MEMS-based Passives for Band-pass Filtering Application," In: *Proceedings of 2011 IEEE TechSym,* Kharagpur, pp. 247–252.

Camargo, E. 1998. *Design of FET Frequency Multipliers and Harmonic Oscillators*, Artech House, Norwood, MA.

Carter, J., Cowen, A., Hardy, B., Mahadevan, R., Stonefield, M. and Wilcenski, S. 2005. *PolyMUMPs Design Handbook,* Revision 11, MEMSCAP Inc., Durham, NC.

Chakraborty, S. and Bhattacharyya, T.K. 2010. "Development of a Surface Micro-machined Binary Logic Inverter for Ultra-low Frequency MEMS Sensor Applications," *Journal of Micromechanics and Microengineering*, 20:105026(15pp).

Coventor. 2008. *MEMS Design and Analysis Tutorials*, vol. 1, Coventor Inc., Cary, NC, 2008.

Faber, M.T., Chramiec, J. and Adamski, M.E. 1995. *Microwave and Millimeter-Wave Diode Frequency Multipliers*, Artech House, Norwood, MA.

Golio, M. 2002. *RF and Microwave Semiconductor Device Handbook*. CRC Press, Boca Raton, FL.

Li, M., Tang, H.X. and Roukes, M.L. 2007. "Ultra-sensitive NEMS-based Cantilevers for Sensing, Scanned probe and Very High-frequency Applications," *Nature Nanotechnology*, 2:114–120.

Lin, Y.-W., Lee, S., Li, S.-S., Xie, Y., Ren, Z. and Nguyen, C.T.-C. 2004. "Series-resonant VHF Micromechanical Resonator Reference Oscillators," *IEEE Journal of Solid-State Circuits*, 39(12): 2477–2491.

Nguyen, C.T.-C. 2007. "MEMS Technology for Timing and Frequency Control," *IEEE Transactions on Ultrasonics, Ferroelectrics and Frequency Control*, 54(2):251–270.

Pandey, A.K. and Pratap, R. 2007. "Effect of Flexural Modes on Squeeze Film Damping in MEMS Cantilever Resonators," *Journal of Micromechanics and Microengineering*, 17:2475–2484.

Polytec. 2012. Website of Polytec GmbH, Waldbronn, *http://www.polytec.com*

Rebeiz, G.M. 2003. *RF MEMS Theory, Design, and Technology*, John Wiley & Sons, Hoboken, New Jersey.

Senturia, S.D. 2001. *Microsystem Design*, Kluwer Academic Publishers.

Wang, H., Nezich, D., Kong, J. and Palacios, T. 2009. "Graphene Frequency Multipliers." *Electron Device Letters*, 30:547–549.

Wang, J., Ren, Z. and Nguyen, C.T.-C. 2004. "1.156-GHz Self-aligned Vibrating Micromechanical Disk Resonator," *IEEE Transactions on Ultrasonics, Ferroelectrics and Frequency Control*, 51:1607–1628.

Wang, Z., Zhang, Z., Xu, H., Ding, L., Wang, S. and Penga, L.-M. 2010. "A High-performance Top-gate Graphene Field-effect Transistor Based Frequency Doubler," *Applied Physics Letters*, 96:173104.

Ward, J., Schlecht, E., Chattopadhyay, G., Maestrini, A., Gill, J., Maiwald, F., Javadi, H. and Mehdi, I. 2004. "Capability of THz Sources Based on Schottky Diode Frequency Multiplier Chains," *IEEE MTT-S International Microwave Symposium Digest*, 3:1587–1590.